\documentclass[preprint,showpacs,titlepage,aps,prd,
amsmath,byrevtex,nofootinbib]{revtex4}

\usepackage{graphicx}
\usepackage[dvips]{epsfig}
\newcommand{\gev}{ ~{\rm GeV} } 
 
\newcommand{\tetatt}{\mbox{$\theta_{23}$}}
\begin{document}

\preprint{FERMILAB-PUB-05-196-T}
\preprint{hep-ph/0505202}
\vspace*{-2cm}

\title{Physics Potential of the Fermilab NuMI beamline}
 
\author{{\large Olga Mena and Stephen Parke}}

\vspace*{1.0cm}
\affiliation{{\sl Theoretical 
Physics Department,
Fermi National 
Accelerator Laboratory 
\\
P.O.Box 500, Batavia, 
IL 60510, USA}\\
omena@fnal.gov 
parke@fnal.gov
}
July 25, 2005

\vglue 1.4cm

\begin{abstract}
We explore the physics potential of the NuMI beamline with a detector located 10 km off-axis at a distant site ($810$ km). We study the sensitivity to $\sin^2 2 \theta_{13}$ and to the CP-violating parameter $\sin \delta$ as well as the determination of the neutrino mass hierarchy by exploiting the $\nu_\mu \to \nu_e$ and $\bar{\nu}_\mu \to \bar{\nu}_e$ appearance channels. The results are illustrated for three different experimental setups to quantify the benefits of increased detector sizes, proton luminosities and $\nu_e$ detection efficiencies.    
\end{abstract}
\maketitle
Neutrino oscillations have been observed and robustly established by the data from solar~\cite{solar,sks}, atmospheric~\cite{SK}, reactor~\cite{kam} and long-baseline neutrino experiments~\cite{k2k}. These results indicate the existence of non-zero neutrino masses and mixings. The new parameters can be accommodated
 via the three neutrino PMNS mixing
 matrix\footnote{We restrict ourselves to a three-family neutrino scenario analysis. The unconfirmed LSND signal cannot be explained in terms of
 neutrino oscillations within this scenario, but might require
 additional light sterile neutrinos or more exotic explanations
 ~\cite{gabriela}.
 The ongoing  MiniBooNE experiment~\cite{miniboone} is expected to explore all
 of the LSND oscillation parameter space~\cite{LSND}.},
 the leptonic analogue to the CKM matrix in the quark sector. Neutrino oscillations within this scenario are described by six parameters: two mass squared differences\footnote{$\Delta m_{ij}^{2} \equiv m_i^2 -m_j^2$ throughout the paper.} ($\Delta m_{21}^2$ and $\Delta m_{32}^2$), three Euler angles ($\theta_{12}$, $ \theta_{23}$ and $\theta_{13}$) and one Dirac CP phase $\delta$. 
The standard way to connect the solar, atmospheric, reactor and accelerator data with the six oscillation parameters listed above is to identify the two mass splittings and the two mixing angles which drive the solar and atmospheric transitions with ($\Delta m_{21}^2$, $\theta_{12}$) and ($|\Delta m_{32}^2|$, $\theta_{23}$), respectively. The sign of the atmospheric mass splitting $\Delta m_{32}^2$ with respect to the solar doublet is one of the unknowns within the neutrino sector, i.e. we do not know if the neutrino mass spectrum is normal ($\Delta m^2_{32}>0$) or inverted ($\Delta m^2_{32}<0$). 
 The best fit point for the combined analysis of solar neutrino data~\cite{snosalt} together with  KamLAND reactor data~\cite{kamland} is at $\Delta m_{21}^2=8.0 \times 10^{-5}$ eV$^2$ and $\sin^{2}\theta_{12}=0.31$\footnote{We use the notation of Ref.~\cite{sinsq} throughout.}. 
The 90\% C.L. allowed ranges of the atmospheric neutrino oscillation
parameters obtained by the Super-Kamiokande experiment are~\cite{SKatm}\footnote{For the numeric analysis presented here, we will use  $|\Delta m^2_{32}| = 2.4 \times 10^{-3} \rm{eV}^2$, which
lies within the best fit values for the Super-Kamiokande~\cite{SKatm}
and K2K~\cite{K2K} experiments.}:
\begin{equation}
|\Delta m^2_{32}|\simeq|\Delta m^2_{31}| =(1.5 - 3.4)\times10^{-3}{\rm eV^2},~~~~
0.36<\sin^2 \theta_{23}< 0.64 
\end{equation}
\indent The mixing angle $\theta_{13}$ (which connects the solar and atmospheric neutrino realms) and the amount of CP violation in the leptonic sector are undetermined. At present, the upper bound on the angle $\theta_{13}$ coming from CHOOZ reactor neutrino data~\cite{chooz} is: 
\begin{eqnarray}
0 \leq & \sin^2 \theta_{13} & < 0.04 
\end{eqnarray}
at the 90 \% confidence level at 
$\Delta m^2_{31} = 2.5 \times 10^{-3} {\rm eV^2}$. 
This constraint depends on the precise value of
$\Delta m^2_{31}$, with a stronger (weaker) constraint at higher
(lower) allowed values of $|\Delta m^2_{31}|$.
Future reactor neutrino oscillation experiments could measure the value of $\sin^{2} \theta_{13}$, as explored in detail in Ref.~\cite{white}. 
Current neutrino oscillation experiments do not have any sensitivity to the
 CP-phase $\delta$. The experimental discovery of the existence of CP violation in the leptonic sector, together with the discovery of the Majorana neutrino character, would point to leptogenesis as the source for the baryon asymmetry of the universe, provided that accidental cancellations are not present. 

The main aim of this paper is a careful study of the sensitivity to the currently unknown parameters mentioned above, that is to  the small mixing angle $\theta_{13}$, to the ordering of the neutrino mass spectrum, and to the amount of CP violation in the leptonic sector, which could be achieved by exploiting the NuMI neutrino beamline. We thus concentrate on the NuMI beam potential exploited in an off-axis configuration as proposed by the NO$\nu$A experiment
~\cite{nova,nova2}. The location of the far detector is at 10 km off-axis with a baseline of 810 km. The mean neutrino energy is 2.3 GeV. We have considered two possible $\nu_e$ ($\bar\nu_e$) detection techniques. First, the possibility of a 30 kton totally active low Z tracking calorimeter detector, as the one considered in the revised NO$\nu$A proposal~\cite{nova2}.  The efficiencies of such a detector for $\nu_e$ ($\bar\nu_e$) identification is approximately $24\%$ and the background is typically two-thirds from electron (anti)neutrinos in the beam produced from muon and kaon decays and one third from neutral current events faking electron neutrinos. An alternative detection method explored here is the one provided by a Liquid Argon TPC detector, as the technique described in the FLARE Letter of Intent~\cite{flare}. The efficiency of such a detector  to identify the $\nu_e$ ($\bar\nu_e$) CC interaction is $80\%$ and the background is dominated by the intrinsic $\nu_e$ and $\bar{\nu}_e$ components of the beam~\cite{flare}. 

The statistics at the far detector is governed by the product of three parameters: the total number of protons on target ($n_p$), the detector mass ($m_D$), and the detector efficiencies to $\nu_e$ ($\bar\nu_e$) identification ($\epsilon$). Here we have studied the physics potential  of three different experimental scenarios:

\begin{itemize}
\item \textbf{Small:} As a first step, we consider a \textbf{``Small''} experimental setup without a Proton Driver. The number of protons on target without a Proton Driver is $6.5\times 10^{20}$ per year. We have considered three and a half years of running in each polarity (i.e. three and a half years of neutrino and three and a half years of antineutrino data taking). Consequently, in this first scenario, the statistical figure of merit $n_p \times m_D \times \epsilon$ is equal to a total of $3.3 \times 10^{22}$ in units of number of protons times kton. This setup could be achieved, for instance, with the 30 kton totally active low Z tracking calorimeter detector at $24\%$ efficiency described above or with a 9 kton Liquid Argon TPC detector at $80\%$ efficiency.
\item  \textbf{Medium:} We study a possible upgrade of the \textbf{Small} configuration, referred to as the \textbf{``Medium''} experimental setup by increasing the statistical figure of merit by a \textbf{factor of five}. 
This statistics factor could be accomplished if the mass of the liquid argon detector is upgraded to 45 kton without a Proton Driver in three and a half years running in each polarity. Or, equivalently, the same statistics could be achieved in the $NO\nu$A experiment running for four and a half years with a Proton Driver in each polarity (i.e. four and a half years of neutrino and four and a half years of antineutrino data taking). With a Proton Driver, the number of protons on target is assumed to be $25.2\times 10^{20}$ per year. Thus the statistics running with a Proton Driver for four and a half years is five times the statistics for running without a Proton Driver for three and a half years~\footnote{This factor of 5 is the ratio of the protons on target per year times the number of years of data taking with and without a proton driver: $\frac{25.2 \times 4.5}{6.5 \times 3.5}$.}.

\item \textbf{Large:} The third experimental setup explored is a \textbf{``Large''} experiment, in which \textbf{both} the initial detector mass and the total number of protons on target are increased by \textbf{a factor of five}. The \textbf{Large} scenario could be obtained, for instance, by the combination of a 45 kt liquid argon detector with a Proton Driver running for four and a half years in both the neutrino and antineutrino modes.
 \end{itemize}

The ratio of the statistics in the three different experimental scenarios \textbf{Small:Medium:Large} is therefore \textbf{1:5:25}.
In the next section we review the neutrino oscillation formalism, while numerical results are presented in the following sections.

\section{PRELIMINARIES}
Since we are exploiting the $\nu_\mu \to \nu_e$ and $\bar{\nu}_\mu \to \bar{\nu}_e$ appearance channels, the observables that we use in our numerical analysis are the number of expected electron neutrino and antineutrino events. For the central values of the already measured oscillation parameters, we have thus computed the expected number of electron and positron events $N_{e^-}^{L, \pm}$ and  $N_{e^+}^{L,\pm}$ at the far detector located $10$ km off-axis at $L= 810$ km, assuming positive or negative hierarchies, which are given by:
\begin{equation}
N^{L,\pm}_{e^{-}(e^{+})}= \int^{E_{max}}_{E_{min}} \; 
dE_{\nu} \; \Phi_{\nu(\bar{\nu})}(E_\nu,L) \; \sigma_{\nu(\bar{\nu})}(E_\nu) \;
 P_{\nu_\mu \nu_e (\bar \nu_\mu \bar \nu_e)}(E_\nu, L, \theta_{13}, \delta, \theta_{23}, \theta_{12}, \pm \Delta m^2_{31}, \Delta m^2_{21})   
\label{eqn:events}
\end{equation}
where $\theta_{23}, \theta_{12}, |\Delta m^2_{31}|$ and $\Delta m^2_{21}$ are taken as perfectly known, $\Phi_{\nu(\bar{\nu})}$ denote the neutrino 
fluxes, and $\sigma_{\nu(\bar{\nu})}$ the cross sections. The neutrino (antineutrino) flux, which peaks at $2.3 \gev$, is integrated over a narrow $1 \gev$ energy window ($E_{min}=1.8 \gev$ and $E_{max}=2.8 \gev$). 

The $\nu_\mu \to \nu_e$ ( $\bar \nu_\mu \to \bar \nu_e$) appearance probabilities in long baseline neutrino
oscillation experiments, assuming the normal mass hierarchy, read~\cite{deg1}:
\begin{eqnarray}
P_{\nu_\mu \nu_e}& = & X_+ \theta^2 +Y_+ \theta \cos(\Delta_{31}+\delta) + P_\odot~, \nonumber \\
P_{\bar\nu_\mu \bar\nu_e}& = & X_-  \theta^2 -Y_-  \theta  \cos(\Delta_{31}-\delta)
+P_\odot  .
\label{eqn:e_appear}
\end{eqnarray}
In the last expressions, $\theta=\sin \theta_{13}$ and the coefficients $X_{\pm}$ and $Y_{\pm}$ are determined by
\begin{eqnarray}
X_{\pm} &=& 4 s^2_{23}
\left\{ \frac{\Delta_{31}\sin({aL \mp \Delta_{31}})}{(aL \mp \Delta_{31})} \right\}^2 , 
\nonumber  \\
Y_{\pm} &=& \pm 2\sqrt{X_\pm P_\odot} =\pm 8 c_{12}s_{12}c_{23}s_{23}
\left\{ \frac{\Delta_{31}\sin({aL \mp \Delta_{31}})}{(aL \mp \Delta_{31})} \right\}
\left\{ \frac{\Delta_{21}\sin{({aL})}}{aL}\right\}~,
\label{Y}\\
P_{\odot} & = & c^2_{23} \sin^2{2\theta_{12}} \left\{ \frac{\Delta_{21}\sin{({aL})}}{aL}\right\}^2~,
\nonumber
\end{eqnarray}
where $\Delta_{ij}  \equiv |\Delta m^2_{ij}| L/4E$, and $a = G_F N_e/\sqrt{2}$ denotes the index of refraction 
in matter, $G_F$ being the Fermi constant and $N_e$ is a constant 
electron number density in the Earth.  
We denote the first, second and third terms in Eqs.~(\ref{eqn:e_appear}) as the atmospheric, interference and solar terms, respectively. When $\theta_{13}$ is relatively large, the probability is dominated by the atmospheric term. Conversely, when 
$\theta_{13}$ is very small, the solar term dominates. The interference term is the only one which contains the CP phase $\delta$, and it is clear from Eqs.~(\ref{eqn:e_appear}) that it is also the only one which differs for neutrinos and antineutrinos besides matter effects.

We can ask ourselves whether it is possible to unambiguously determine $\theta_{13}$ and $\delta$ by measuring the transition probabilities $\nu_\mu \to \nu_e$ and $\bar\nu_\mu \to \bar\nu_e$ at fixed neutrino energy and at just one baseline. The answer is no. At fixed neutrino energy $E_\nu$ and  baseline $L$, if ($\theta_{13}$, $\delta$) are the values chosen by Nature, the conditions:
\begin{center}
\begin{equation}
\left.\begin{matrix}
P_{\nu_\mu \nu_e} (\theta^{'}_{13}, \delta^{'}) = P_{\nu_\mu \nu_e}
(\theta_{13}, \delta)\nonumber \cr 
P_{\bar \nu_\mu \bar \nu_e} (\theta^{'}_{13}, \delta^{'}) = P_{\bar \nu_\mu
\bar \nu_e} (\theta_{13}, \delta)
\end{matrix}
\right \}
\label{eqn:binprob}
\end{equation}
\end{center}
can be generically satisfied by another set ($\theta^{'}_{13}$, $\delta^{'}$), known as the \emph{intrinsic degeneracy}~\cite{deg1}. It has also been pointed out that other fake solutions might appear from  unresolved degeneracies in two other oscillation parameters:
\begin{enumerate}
\item The sign of the atmospheric mass difference $\Delta m_{32}^2$ may remain unknown. In this particular case, $P (\theta^{'}_{13}, \delta^{'}, -\Delta m_{32}^{2}) = P (\theta_{13}, \delta, \Delta m_{32}^2)$~\cite{deg2,deg3}. More specifically, the $\nu_\mu \to \nu_e$ ( $\bar \nu_\mu \to \bar \nu_e$) appearance probabilities for the inverted hierarchy are  
\begin{eqnarray}
 P_{\nu_\mu \nu_e}& = & X_-  \theta^2 +Y_-  \theta \cos(\Delta_{31}-\delta) +P_\odot
\nonumber \\
 P_{\bar\nu_\mu \bar\nu_e}& = & X_+  \theta^2 -Y_+  \theta \cos(\Delta_{31}+\delta)
+P_\odot.
\end{eqnarray}
\item Disappearance experiments only give us information on $\sin^{2} 2 \theta_{23}$: is $\theta_{23}$ in the first octant, $0<\theta_{23}<\pi/4$, or is it in the second one, $\theta_{23}\to \pi/2 -\theta_{23}$?. In terms of oscillation probabilities, $P (\theta^{'}_{13}, \delta^{'}, \frac{\pi}{2}-\tetatt) = P (\theta_{13}, \delta ,\tetatt)$~\cite{deg2,deg4}.
\end{enumerate}
\begin{figure}[t]
\begin{center}
\begin{tabular}{ll}
\hskip -0.5cm
\epsfig{file=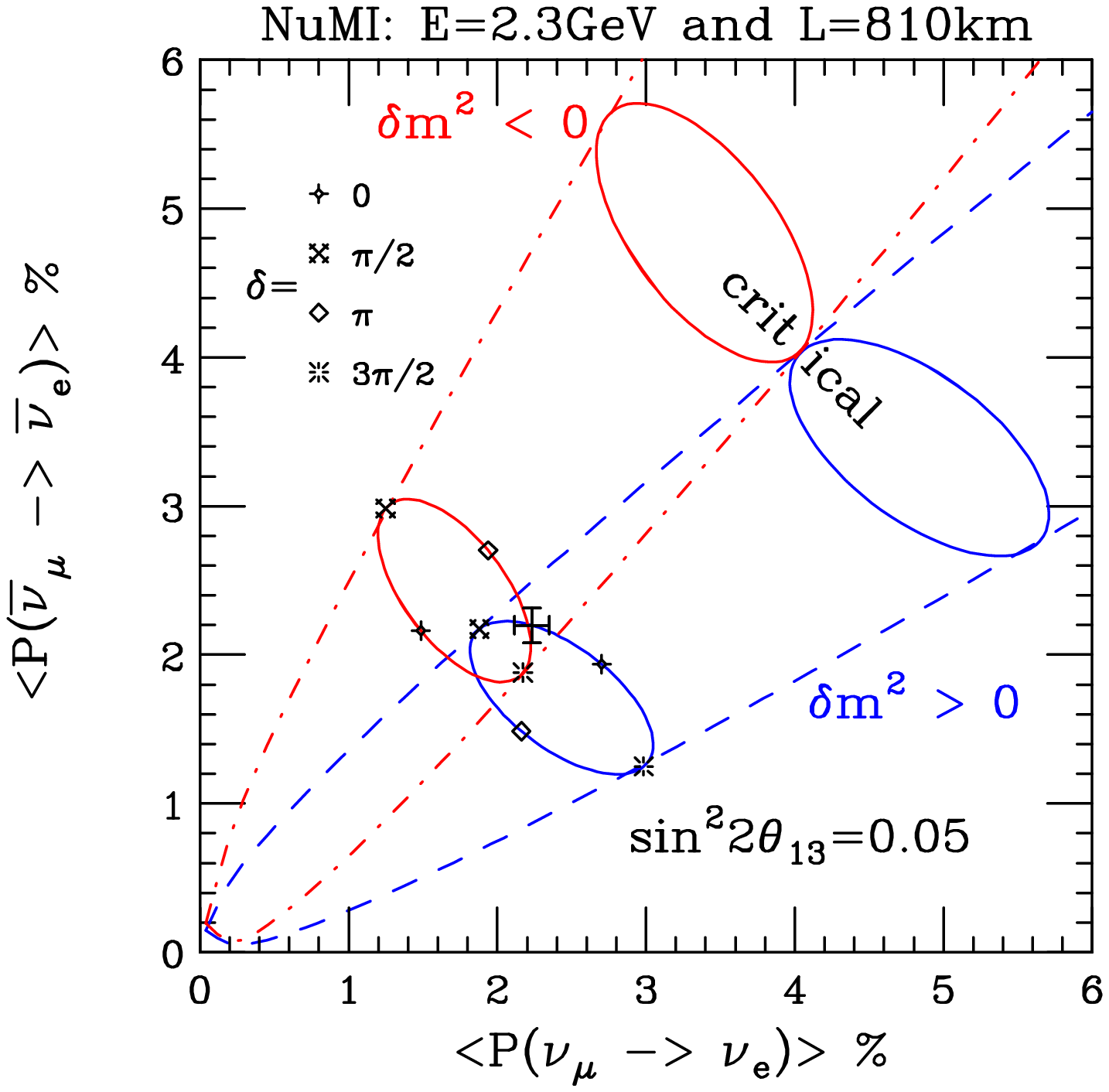, width=8.1cm} &
\hskip 0.cm
\epsfig{file=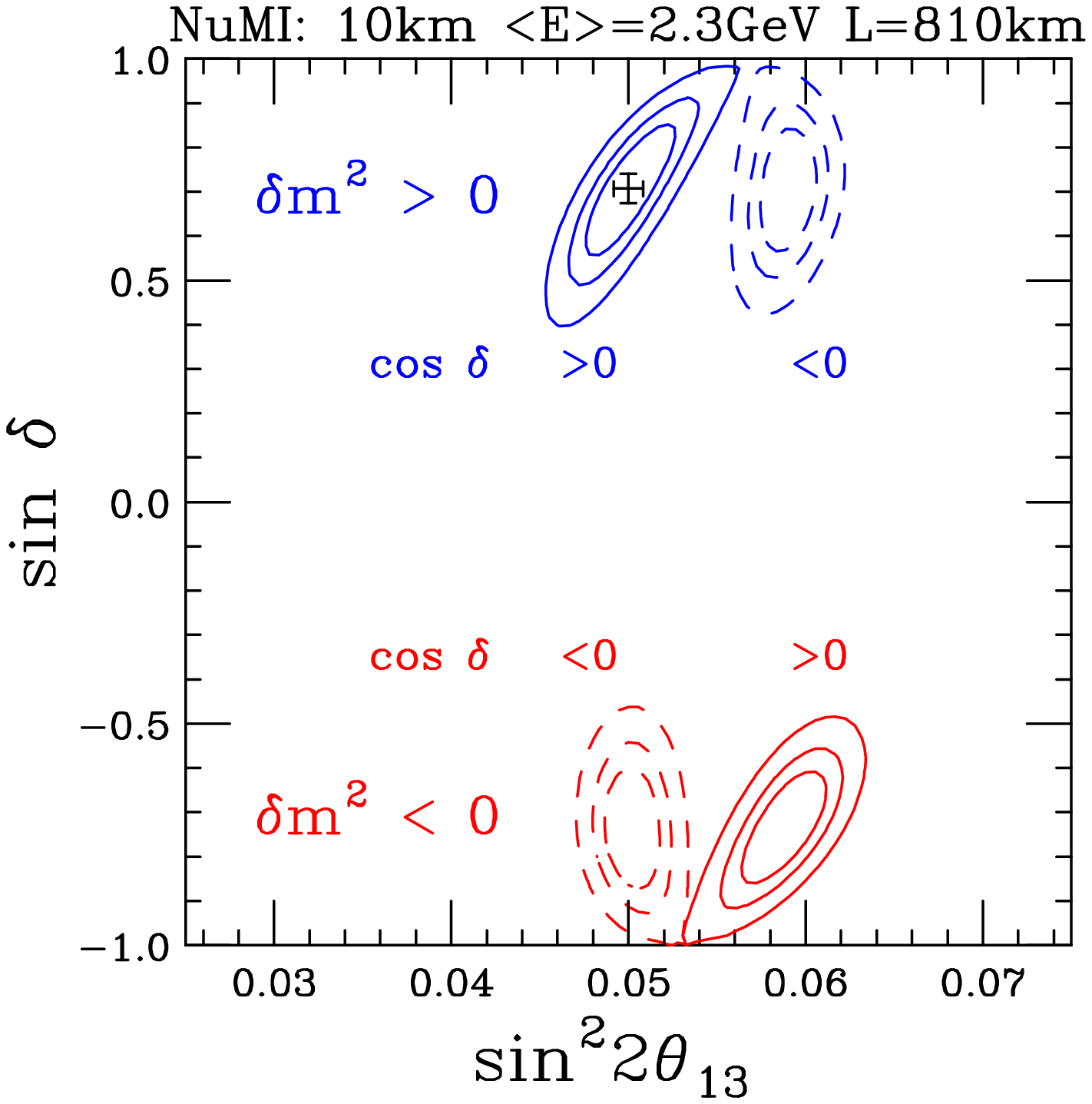, width=8.1cm} \\
\hskip 3.2truecm
{\small (a)}            &
\hskip 3.8truecm
{\small (b)}
\end{tabular}
\end{center}
\caption{\textit{(From ref.~\cite{mp2})(a) The bi-probability diagram for NuMI 10 km off-axis showing the allowed regions for both the normal (dashed) and inverted (dot-dashed) hierarchies as well as the ellipses for $\sin^2 2\theta_{13} =0.05$.
The large ``{\large $+$}'' marks the neutrino and anti-neutrino probabilities with the CP phase, $\delta = \pi/4$, assuming the normal hierarchy. The ellipses and point along the diagonal
labeled critical correspond to the largest values for which there is overlap between the normal and inverted hierarchies. (b)The allowed regions in the $\sin^2 2\theta_{13}$ v $\sin \delta$ plane for the NuMI 10 km off-axis
experiment, assuming the true solution is the normal hierarchy with $\sin^2 2\theta_{13} =0.05$ and $\delta = \pi/4$ (``{\large $+$}''). The upper blue
(lower red) contours are for the normal (inverted) hierarchy
whereas the solid (dashed) contours are for $\cos \delta >0$ $(<0)$. The experimental setup corresponds to the \textbf{Large} scenario considered here. The ellipses correspond to 68, 90 and 99\% C.L. contours.}}
\label{fig:prob}
\end{figure}

Extensive work has been devoted recently to eliminate such fake solutions. In simple terms, when considering data from two or more experiments, degenerate solutions may occur at different locations in parameter space for different experiments, and therefore could be excluded~\cite{deg1,mnpx,mp2,prevst}.
As shown in Ref.~\cite{mp2}, there exists a simple way to understand if the various degeneracies are eliminated when several experiments are combined.  We briefly review here the analysis of Ref.~\cite{mp2}, and illustrate its results for the NuMI beamline exploited in an off-axis mode. In the overlap region of the $\nu_\mu \to \nu_e$ bi-probability diagram~\cite{MNP}, see Fig.~(\ref{fig:prob})(a), there exist generically four solutions\footnote{We assume $\theta_{23}=\pi/4$; we will show that small variations of this parameter do not affect the results presented here in a significant way.}  for the unknown oscillation parameters $\theta_{13}$ and $\delta$. Two solutions correspond to the normal hierarchy~\cite{deg1} and have approximately equal values of $\sin \delta$ but different values of the sign of $\cos \delta$. The other two solutions are for the inverted hierarchy~\cite{deg2,deg3}, and they also have approximately equal values of $\sin \delta$. This value of $\sin \delta$ for the inverted hierarchy is different than the value of $\sin \delta$ for the normal hierarchical case.
In Ref.~\cite{mp2}, an identity connecting the difference between the mean values of $\sin \delta$ (which governs the amount of leptonic CP violation) for the two hierarchies, to the mean values of $\theta_{13}$ for both hierarchies, is derived. Such an identity turns out to be extremely helpful in understanding if the combination of several experiments can eliminate the fake solutions, since the location of the fake solutions in the ($\sin^{2} 2 \theta_{13}$, $\sin \delta$) plane can be computed in a straightforward manner.  If we apply this identity relating the solutions corresponding to the positive and negative hierarchy to the NuMI 10 km off-axis experiment, it was found: 
\begin{eqnarray}
 \langle \sin \delta \rangle_+
- \langle \sin \delta \rangle_- 
& = &  1.41 \sqrt{\sin^2 2\theta_{13} \over 0.05}~,  
\label{eqn:difsins}
\end{eqnarray}
where  $\langle \sin \delta \rangle_{+(-)}$ are the mean values of the two solutions of $\sin \delta$ for each hierarchy, see Ref.~\cite{mp2} for details\footnote{If the detector is located 12 km off-axis at the far site, the numerical factor in Eq.~(\ref{eqn:difsins}) is 1.46 instead of 1.41. Consequently, the changes associated with placing the detector at 12 km off-axis rather than at 10 km off-axis are small.}. For the sake of illustration we show in Fig.~(\ref{fig:prob})(b) the $\chi^2$ contours for NuMI 10 km off-axis in the \textbf{Large} experimental setup explored in the present study, assuming that the true solution is the normal hierarchy and that the values of 
($\sin^2 2 \theta_{13}$, $\delta$) are (0.05, $\pi/4$), respectively. The NuMI 10 km off-axis is operated above oscillation maximum: there are thus four solutions in  the $(\sin^2 2\theta_{13},~\sin \delta)$ plane.

The existence of such a simple relation, Eq.~(\ref{eqn:difsins}), among the true and fake solutions in terms of $\sin \delta$, together with the fact that it is precisely $\sin \delta$ the quantity which drives the amount of leptonic CP-violation, has motivated us to consider $\sin \delta$ as the relevant parameter in our analytical and numerical studies. In the case of the T2K experiment, the difference between the true and fake solutions for the CP violating parameter $\sin \delta$ is $0.47$ at $\sin^2 2 \theta_{13}=0.05$. This  factor of 3 decrease with respect to the NuMI off-axis experiment is primarily due to the T2K baseline is $1/3$ the NuMI off-axis baseline.

If the mixing angle $\theta_{23}\neq \pi/4$, the ($\sin^{2} 2 \theta_{13}$, $\sin \delta$) plane should be translated into the ($2 \sin^2 \theta_{23} \sin^2 2\theta_{13}$, $\sqrt{2} \cos \theta_{23} \sin \delta$) plane, as shown in Ref.~(\cite{mp2}). Assuming this \emph{mapping}, the results presented in the next sections would be almost identical even if $\theta_{23}\neq \pi/4$.

\section{$\sin^{2} 2 \theta_{13}$ sensitivity}
\begin{figure}[t]
\begin{center}
\epsfig{file=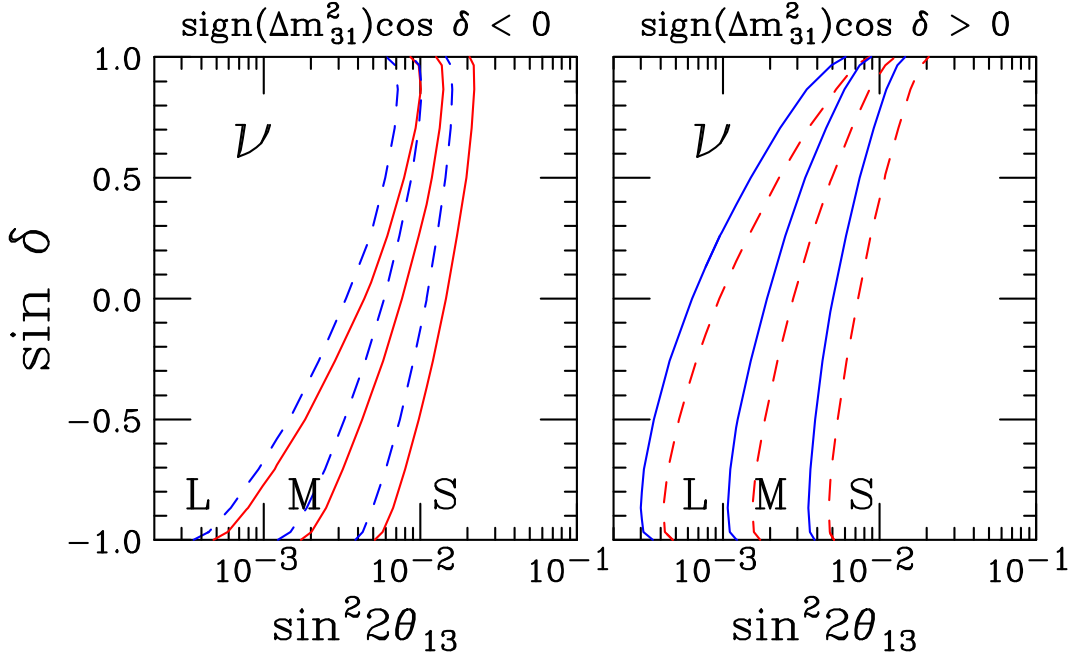, width=14.1cm} 
\end{center}
\caption{\textit{The $99\%$ CL sensitivity contours in the ($\sin^{2} 2 \theta_{13}$, $\sin \delta$) plane for $\sin^{2} 2 \theta_{13}$, exploiting only the data in the neutrino channel. In the left panel we depict the sensitivities for the most restrictive choice of parameters, which corresponds to $\cos\delta<0$ and normal hierarchy  (dashed blue curve) or to $\cos\delta>0$ and negative hierarchy (solid red curve). In the right panel we show the sensitivity for the choice of parameters where $\cos\delta>0$ in  the normal hierarchy (solid blue curve) or $\cos\delta<0$ and the hierarchy is inverted (dashed red curve). The labels \textbf{L}, \textbf{M} and \textbf{S} correspond to the \textbf{Large}, \textbf{Medium} and \textbf{Small} experimental setups explored in this study, respectively. See Table (\ref{tab:th13sens}) in the Appendix for numerical limits.}}

\label{fig:sensnu}
\end{figure}

\begin{figure}[h]
\begin{center}
\epsfig{file=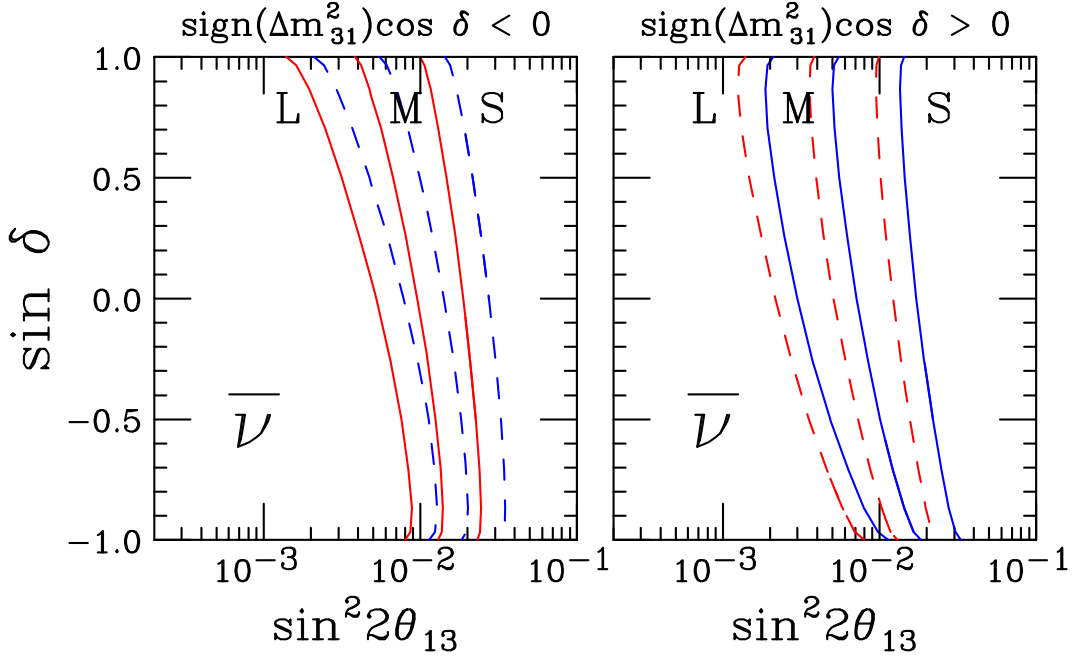, width=14.1cm} 
\end{center}
\caption{\textit{The $99\%$ CL sensitivity contours in the ($\sin^{2} 2 \theta_{13}$, $\sin \delta$) plane for $\sin^{2} 2 \theta_{13}$, exploiting only the data in the antineutrino channel. The curves and labels are same as those of Fig.~(\ref{fig:sensnu}).}} 
\label{fig:sensanu}
\end{figure}

\begin{figure}[t]
\begin{center}
\epsfig{file=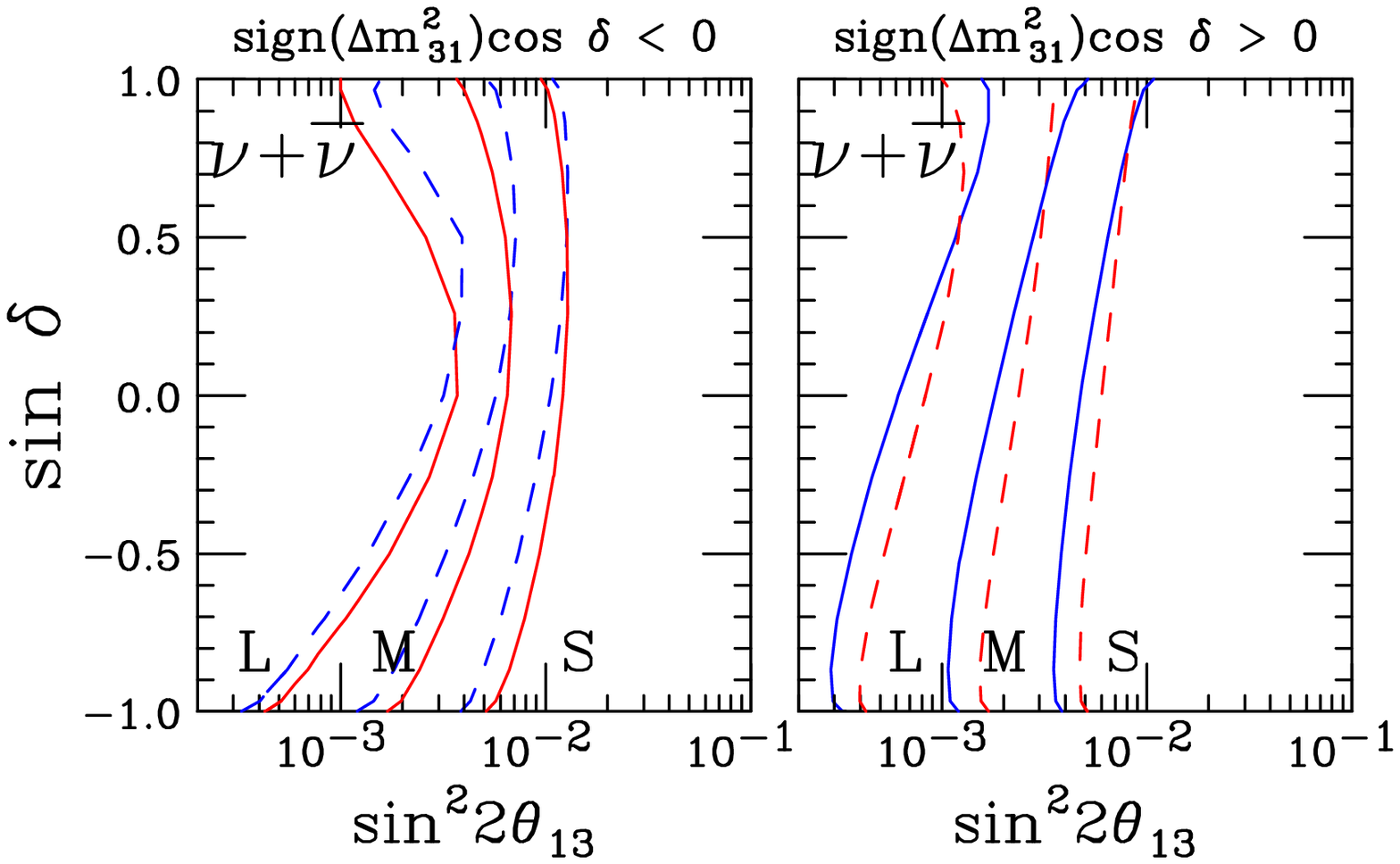, width=14.1cm} 
\end{center}
\caption{\textit{The $99\%$ CL sensitivity contours in the ($\sin^{2} 2 \theta_{13}$, $\sin \delta$) plane for $\sin^{2} 2 \theta_{13}$, exploiting both the data in the neutrino and in the antineutrino channel.  The curves and labels are same as those of Fig.~(\ref{fig:sensnu}).}}
\label{fig:sensnuanu}
\end{figure}

In the present section we explore the sensitivity of the proposed NuMI long baseline off-axis experiment to a $\nu_\mu \to \nu_e$
($\bar{\nu}_\mu \to \bar{\nu}_e$) oscillation search in the appearance mode.
In Figs.~(\ref{fig:sensnu}), (\ref{fig:sensanu}) and (\ref{fig:sensnuanu}) we depict the sensitivity contours in the ($\sin^{2} 2 \theta_{13}$, $\sin \delta$) plane for the three different setups described in the introduction for neutrino data, antineutrino data and for a combined analysis of both neutrino and antineutrino data, respectively.
  
We have four different curves for neutrinos (for antineutrinos as well as for the combination of the two channels), according to the mass spectrum hierarchy  and to the sign of $\cos \delta$. 
Since both the sign of the atmospheric mass splitting and the sign of the $\cos \delta$ may remain unknown, the maximum sensitivity to $\sin^2 2 \theta_{13}$ versus $\sin \delta$ in a given setup should be identified with the most conservative curve among the four possibilities. We describe in detail Fig.~(\ref{fig:sensnu}) but the same criterion should be applied to Figs.~(\ref{fig:sensanu}) and (\ref{fig:sensnuanu}). In the  left panel of Fig.~(\ref{fig:sensnu}) it is depicted the sensitivity to $\sin^2 2 \theta_{13}$ versus the CP violating quantity $\sin \delta$ in the two conservative pictures, i.e. when the sign of $\cos \delta<0$ in the normal hierarchical scenario or when $\cos \delta>0$ in the inverted hierarchy picture. Being the sign of $\cos\delta$ and the neutrino mass hierarchy unknowns within the neutrino mixing sector, the sensitivity to $\sin^2 \theta_{13}$ must be associated with the tightest bound among the two possibilities.
In the particular case that we are describing here, i.e, the one exploiting the neutrino channel information, the sensitivity curve is given by the red solid curve in the left panel of Fig.~(\ref{fig:sensnu}). In the right panel of Fig.~(\ref{fig:sensnu}) we show the most optimistic scenarios where $\cos\delta>0$ in the normal hierarchy picture or $\cos\delta<0$ in the inverted one. However, we should remark here that these curves do not represent the true sensitivity to $\sin^2 2 \theta_{13}$, which is the first priority of the near future neutrino oscillation experiments, before the measurements of the sign of the atmospheric mass difference and the sign of $\cos\delta$. 

\begin{figure}[t]
\begin{center}
\begin{tabular}{ll}
\hskip -0.5cm
\epsfig{file=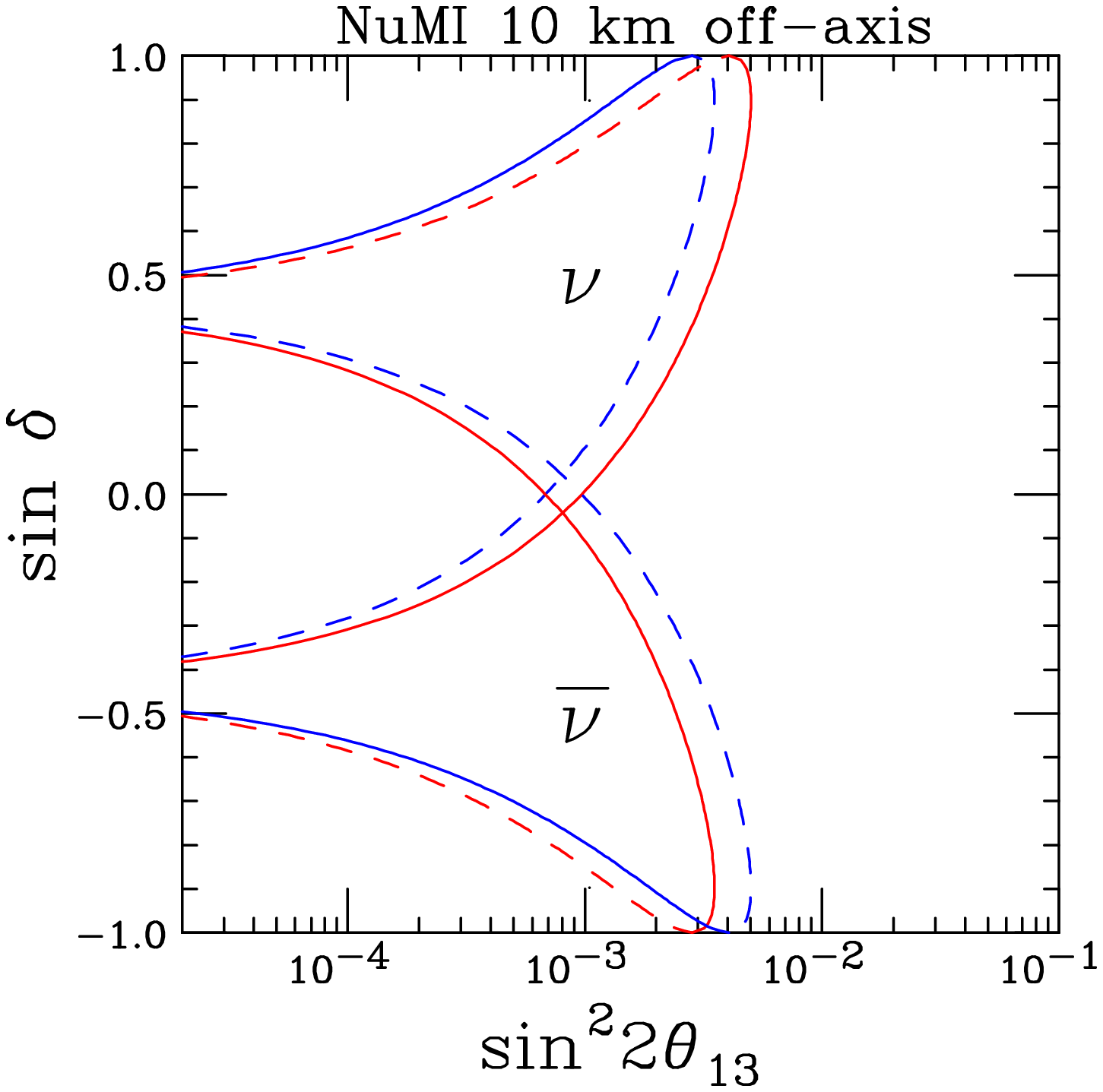, width=8.1cm} &
\hskip 0.cm
\epsfig{file=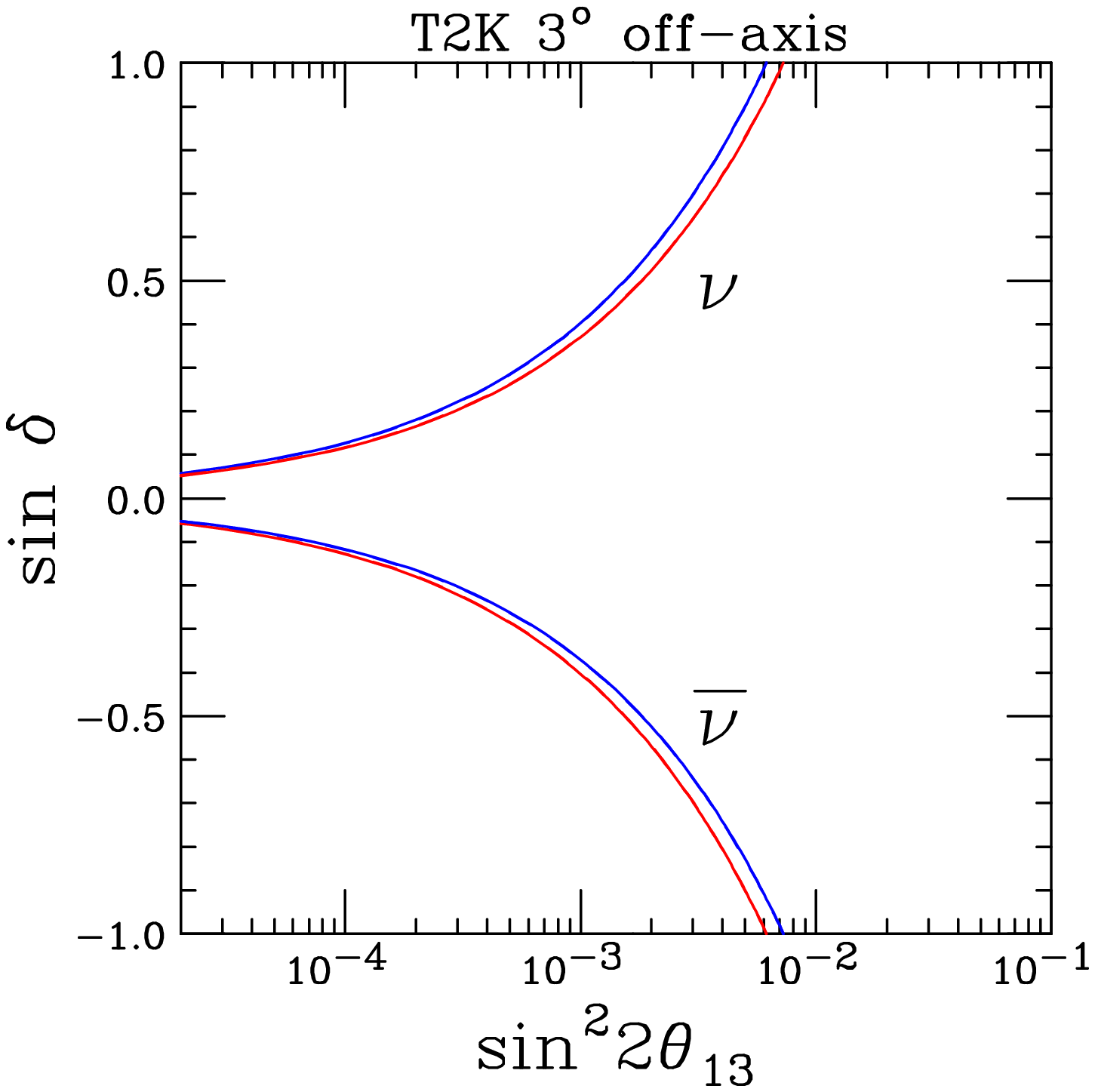, width=8.1cm} \\
\hskip 3.2truecm
{\small (a)}            &
\hskip 3.8truecm
{\small (b)}
\end{tabular}
\end{center}
\caption{\textit{(a) Zero-mimicking solution for the NuMI off-axis experiment, in the ($\sin^{2} 2 \theta_{13}$, $\sin \delta$) plane, for neutrinos (upper curves) and antineutrinos (lower curves). The blue (red) depicts normal (inverted) hierarchy. The vacuum zero-mimicking solutions (not depicted here) are located between the two hierarchies, therefore matter effects are small. The solid (dashed) curves depict the positive (negative) sign for $\cos \delta$. (b) The same as in (a) but for the T2K experiment, as an illustration of the zero-mimicking solutions at vacuum oscillation maximum, $\Delta_{31}=\frac{\pi}{2}$.}}
\label{fig:mimic}
\end{figure}

We point out here the existence of \emph{zero-mimicking solutions}: there exists, at a fixed neutrino energy and baseline, a point in 
the  ($\sin^{2} 2 \theta_{13}$, $\sin \delta$) plane at which the first (atmospheric) and second (interference) terms in Eq.~(\ref{eqn:e_appear}) exactly cancel, and the situation is indistinguishable from the one in which $\sin\theta_{13}=0$. In vacuum, the location of the \emph{zero-mimicking solution} at a given $\delta$ is  
\begin{equation}
\sin 2 \theta_{13} =  - 2 \frac{\Delta_{21}}{\sin\Delta_{31}} \frac{\sin 2 \theta_{12}}{\tan \theta_{23}} \cos (\Delta_{32} \pm \delta)~,
\end{equation}
where the sign +(-) refers to neutrinos (antineutrinos). If the experiment were operating at the vacuum oscillation maximum, the zero-mimicking solution would be located at:
\begin{equation}
\sin 2 \theta_{13} =  \pm 2 \ \Delta_{21} \frac{\sin 2 \theta_{12}}{\tan \theta_{23}} \sin \delta~,
\end{equation}
where the sign +(-) refers to neutrinos (antineutrinos). At the vacuum oscillation maximum the  $\sin \delta$ value of the zero-mimicking solution  is positive for neutrinos (negative for antineutrinos). Figure (\ref{fig:mimic})(b) depicts the zero-mimicking solution for the T2K experiment~\cite{t2k}. T2K will use a steerable neutrino beam from JHF to Super-Kamiokande and/or Hyper-Kamiokande as the far detector(s). The mean energy of the neutrino beam will be tuned to be at vacuum oscillation maximum, $\Delta_{31}= \frac{\pi}{2}$, which implies a mean neutrino energy $\langle E_\nu \rangle =0.6$ GeV at the baseline of  295 km, using $|\Delta m^2_{31} |= 2.4 \times 10^{-3}$eV$^2$. This neutrino energy can be obtained with a 3$^o$ off-axis beam. Since T2K will be operated at vacuum oscillation maximum, this experiment is insensitive to the CP conserving quantity $\cos \delta$.

When matter effects are considered, the situation is slightly more complicated. The zero-mimicking solution is in general located at different points in the ($\sin^{2} 2 \theta_{13}$, $\sin \delta$) plane, as illustrated in Fig.~(\ref{fig:mimic})(a), showing the zero-mimicking solution for the NuMI 10 km off-axis experiment\footnote{The NuMI beam energy (2.3 GeV) is about 30\%
above the vacuum oscillation maximum energy for its baseline, i.e 810 km. Since NuMI 10 km off-axis is operated above oscillation maximum, this experiment is sensitive to the sign of $\cos \delta$.}:
\begin{equation}
\sin 2 \theta_{13} =  - 2 \frac{\sin 2 \theta_{12}}{\tan \theta_{23}}
\left\{ \frac{\Delta_{21}\sin{({aL})}}{aL}\right\}
\left\{ \frac{(aL \mp \Delta_{31})}{\Delta_{31}\sin({aL \mp \Delta_{31}})} \right\}
\cos (\Delta_{32} \pm \delta)~,
\end{equation}

The existence of zero-mimicking solutions allow us to understand the shape of the sensitivity curves, i.e. the Figs.~(\ref{fig:sensnu}), (\ref{fig:sensanu}) and (\ref{fig:sensnuanu}).
Exploiting the neutrino data, the sensitivities will therefore improve enormously as the experimental setup (i.e. the statistics) is improved, as long as $\sin \delta$ is negative. The sensitivity to $\sin^2 2 \theta_{13}$ is maximal when $\sin \delta = -1$. On the other hand, if the antineutrino data is exploited, the situation is reversed: the sensitivities will improve in a significant way as the setup is upgraded if $\sin \delta$ is positive, and the optimal sensitivity in this case is reached when $\sin \delta =1$.
When combining the data from the neutrino and antineutrino channels, the sensitivity curves are flatter than in each separate case (i.e. considering only the neutrino data or only the antineutrino one), see Fig.~(\ref{fig:sensnuanu}). This flattening effect on the sensitivity curves when adding the information of both channels increases as the exposure decreases: in the \textbf{Small} experimental setup considered here the curves are basically flat in the ($\sin^{2} 2 \theta_{13}$, $\sin \delta$) plane, see Fig.~(\ref{fig:sensnuanu}).

\section{Mass Hierarchy determination} 

\begin{figure}[t]
\begin{center}
\epsfig{file=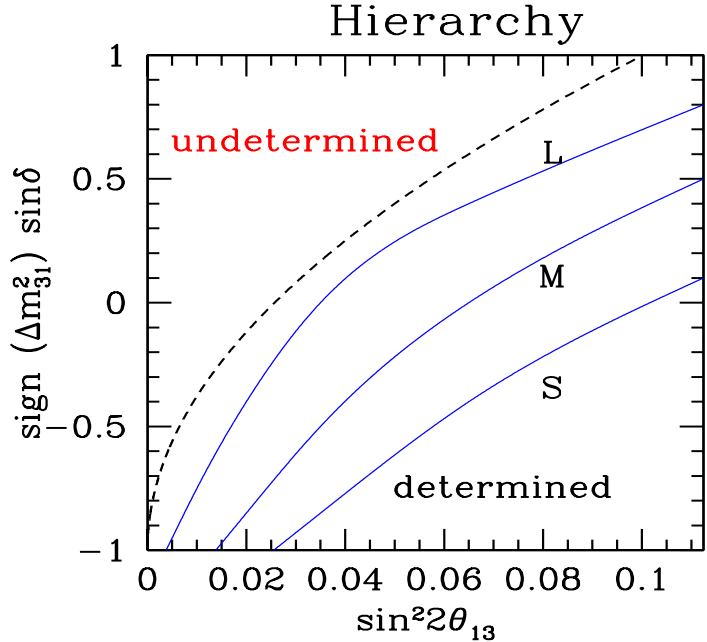, width=10.1cm,height=9.cm} 
\caption{\textit{Sensitivity to the sign($\Delta m^{2}_{31}$)-extraction at the $95\%$ CL within the three reference setups explored in the present study. The labels \textbf{L}, \textbf{M} and \textbf{S} correspond to the \textbf{Large}, \textbf{Medium} and \textbf{Small} experimental setups explored in this study, respectively. The dashed black curve is obtained from Eq.~(\ref{eqn:difsins}) setting $\langle \sin \delta \rangle_- = -1$ ($\langle \sin \delta \rangle_+ = +1$) for the normal (inverted) hierarchy. This is the bound that would be obtained with infinite statistics and in the absence of backgrounds.}}
\label{fig:mass}
\end{center}
\end{figure}

In this section we study the possible extraction of the sign of the atmospheric mass splitting with the NuMI 10 km off-axis experiment, within the context of the three reference experimental setups considered in the present paper. 
We have performed a $\chi^{2}$ analysis of the data in the ($\sin^{2} 2 \theta_{13}$, $\sin \delta$) plane. We assume Nature has chosen the normal or inverted hierarchy and we attempt to fit the data to the expected number of events for the opposite hierarchy. Generically one expects two fake solutions associated with the wrong choice of the hierarchy at fixed neutrino energy and baseline. The $\chi^2$ function, from the combination of the neutrino and antineutrino channels, reads:
\begin{equation}
\chi^2 = \sum_{p = e^+,e^-}  \left(\frac{ {\mathcal N}^{+}_{p} \, - \, 
N^{-}_{p}- {\mathcal B}_{p}}{\delta {\mathcal N}^{+}_{p}}\right)^2~,
\label{eqn:chi2}
\end{equation}
where $\delta {\cal N}^{+}_{p}$ is the statistical error on ${\cal N}^{+}_{p}$, the simulated data
\begin{equation} 
{\mathcal N}^{+}_{p}= {\rm Smear} (N^{+}_{p} + {\mathcal B}_{p})~, 
\label{eqn:data}
\end{equation}
where  ${\cal B}_{p}$ is the number of background events when running in a fixed polarity $p$ and we have performed a Poisson smearing to mimic the statistical uncertainty. 
In Fig.~(\ref{fig:mass}) we depict the results for the sign($\Delta m^{2}_{31}$)-extraction by exploiting the neutrino and antineutrino data in the three reference setups. As expected, the best sensitivity is reached with the most ambitious scenario, relying on the proton driver option. The shape of the exclusion lines can be easily understood in terms of matter effects, which are quite significant for the NuMI off-axis experiment as can be clearly noticed from the bi-probability diagram, Fig.~(\ref{fig:prob})(a). The shift observed in the bi-probability events is proportional to the size of the matter effects, which are obviously crucial to resolve the hierarchy of the neutrino mass spectrum\footnote{Recently, new approaches for determining the type of hierarchy have been proposed~\cite{hieratm} by exploiting other neutrino oscillations channels, such as muon neutrino disappearance, and require very precise neutrino oscillation measurements.}. The sensitivity to the measurement of the sign of the atmospheric mass difference is expected to be better when the sign of $\sin\delta$ is negative: in the case of the \textbf{Medium} experimental setup, the sensitivity to the sign ($\Delta m^2_{31}$)-extraction is lost for positive values of $\sin\delta$. We show as well in Fig.~(\ref{fig:mass}) the theoretical limit on the sign($\Delta m^{2}_{31}$)-extraction, which acts as a rigorous upper bound on the experimental sensitivity curves.  
A possible way to resolve the fake solutions associated to the sign of the atmospheric mass difference would be to combine the data from the proposed NuMI 10 km off-axis and T2K experiments~\cite{mnpx,mp2}. The complementarity of the NuMI and T2K experiments can be  explicitly shown by exploiting the identity given in the introductory Section by Eq.~(\ref{eqn:difsins}) and in Ref.~\cite{mp2}. The difference in the location of the fake solutions associated to the wrong hierarchy in the ($\sin^2 2 \theta_{13}$, $\sin \delta$) plane for these two experiments reads:

\begin{eqnarray}
| ~\langle \sin \delta \rangle_{fake}^{T2K}
- \langle \sin \delta \rangle_{fake}^{NuMI}~|
& = &  0.94 \sqrt{\sin^2 2\theta_{13} \over 0.05} .
\end{eqnarray}

This relation implies that the wrong solutions would appear in different regions of the parameter space for the two experiments, and therefore the fake solutions could be eliminated by a combined NuMI/T2K analysis.\footnote{Based on previous work~\cite{mnpx}, the authors of Ref.~\cite{short} have recently shown that it would be possible to extract the sign of the atmospheric mass difference by exploiting the NuMI off-axis beamline with just a neutrino run, provided that two detectors would be placed at the same $E/L$. The experimental picture would be given by a near detector located before the NO$\nu$A far site (probably at $200$ km, to optimize the sensitivity) and a second detector at the far site.}

\section{CP-$\sin\delta$ measurement} 

\begin{figure}[h]
\begin{center}
\epsfig{file=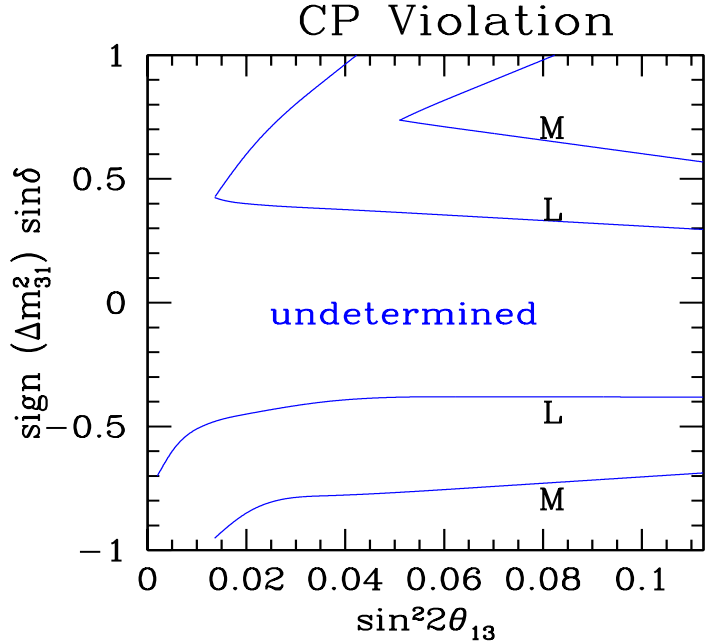, width=10.1cm,height=9.cm} 
\caption{\textit{Sensitivity to CP violation (i.e. $\sin\delta\neq0$) at the $95\%$ CL. The difference in the shape of the two top exclusion curves is due to the additional restrictions coming from the presence of the sign($\Delta m^{2}_{31}$)-degeneracies. If the hierarchy were determined, the top curves would be close to a reflection of the bottom ones. The labels \textbf{L} and \textbf{M} correspond to the \textbf{Large} and \textbf{Medium} experimental setups explored in this study, respectively.}}
\label{fig:cp}
\end{center}
\end{figure}

In the present section we explore the sensitivity to CP violation for the three different experimental scenarios under consideration. The results are summarized in Fig.~(\ref{fig:cp}), in which we depict the exclusion contours at the $95\%$ CL corresponding to the \textbf{Large} and to the \textbf{Medium} setups. The measurement of leptonic CP violation is certainly not within reach for the less ambitious scenario, i.e. the \textbf{Small} experimental setup described in the introductory section, not shown in Fig.~(\ref{fig:cp}).

The exclusion lines depict the value of $\sin\delta$ at which the $95\%$ CL error in the CP violating parameter $\sin \delta$ reaches $\sin \delta=0$, corresponding to the CP conserving case. An important point to note here is that, in order to compute the curves in Fig.~(\ref{fig:cp}), we have considered the impact of the degeneracies associated to the sign of the atmospheric mass difference. It may happen that the error on the fake-sign solutions reaches $\sin \delta=0$ and therefore the analysis would be consistent with CP conservation even if the error in the true solution has not reached $\sin \delta=0$. The mass hierarchy-sign degeneracies affect the CP-sensitivity contours only if sign$(\Delta m^2_{31})\sin \delta$ is positive, as would be expected from the analysis performed in the previous section to the sensitivity to the sign of the atmospheric mass difference.

\section{Analysis in energy bins} 

In the present section we restrict the analysis to the intrinsic degeneracies: we assume that Nature has chosen $\Delta m^2_{31}>0$ and $\theta_{23}=\pi/4$, and we concentrate on only one scenario, the \textbf{Large} setup.
 
In order to eliminate the fake solutions associated with the wrong choice of the sign of $\cos\delta$, we exploit here the energy dependence of the signal. We have thus divided the total number of events in two bins of equal width $\Delta E_\nu=0.5\gev$ \footnote{A very conservative estimate for the neutrino energy resolution in the range of interest is $\Delta E_\nu /E_\nu \sim 50 \%$}. 
The $\chi^2$ function of Eq.~(\ref{eqn:chi2}) at the fixed baseline $L=810$ km now reads:
\begin{equation}
\chi^2  = \sum_{p = e^+,e^-} \sum_i 
\left(\frac{ {\mathcal N}^{+}_{i,p} \, - \, 
N^{+}_{i,p}- {\mathcal B}_{p}}{\delta {\mathcal N}^{+}_{i,p}}\right)^2 \, ,
\label{eqn:chi2bin}
\end{equation}

\begin{figure}[t]
\begin{center}
\begin{tabular}{ll}
\hskip -0.5cm
\epsfig{file=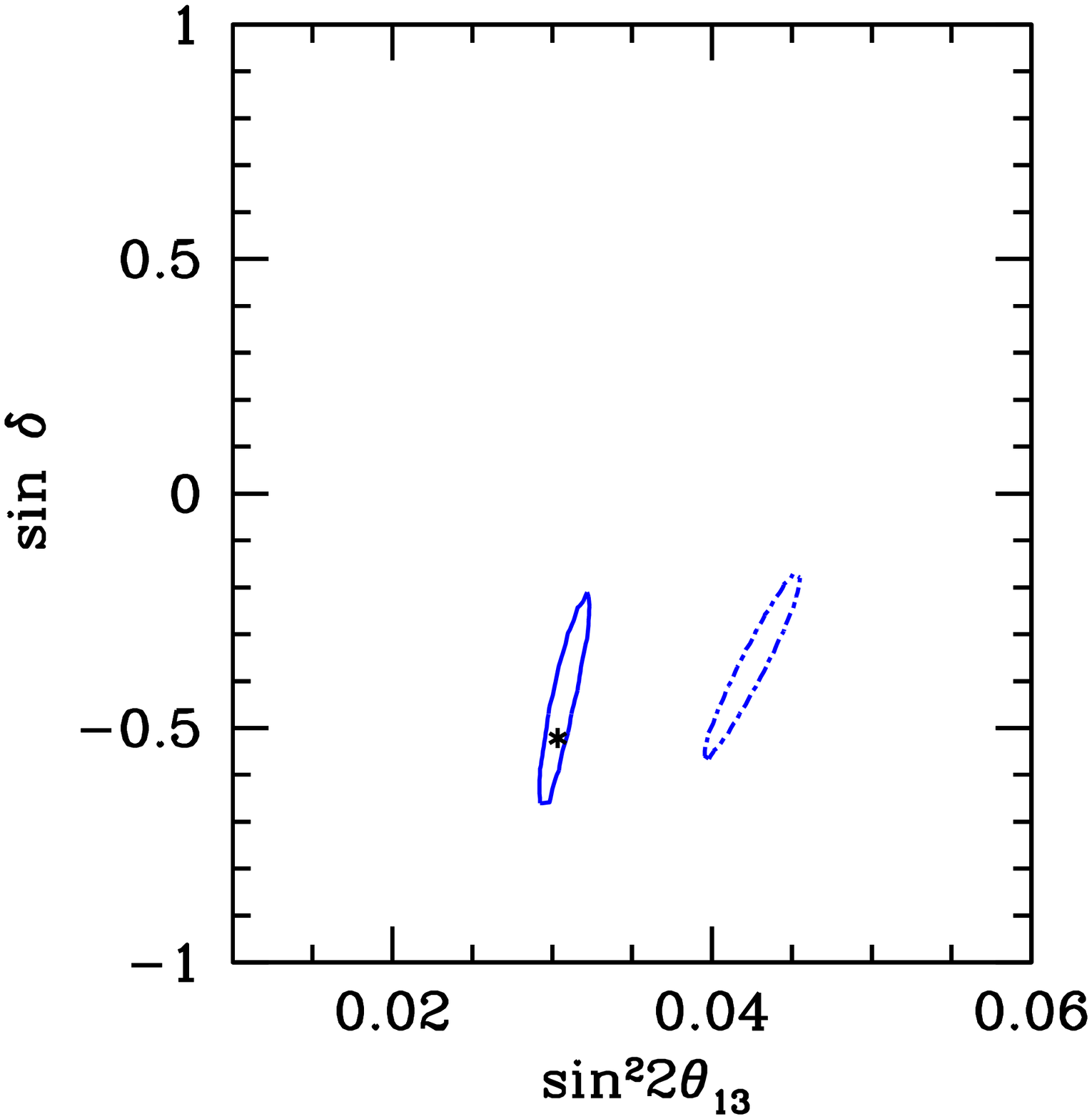, width=9.1cm} &
\hskip -0.5cm
\epsfig{file=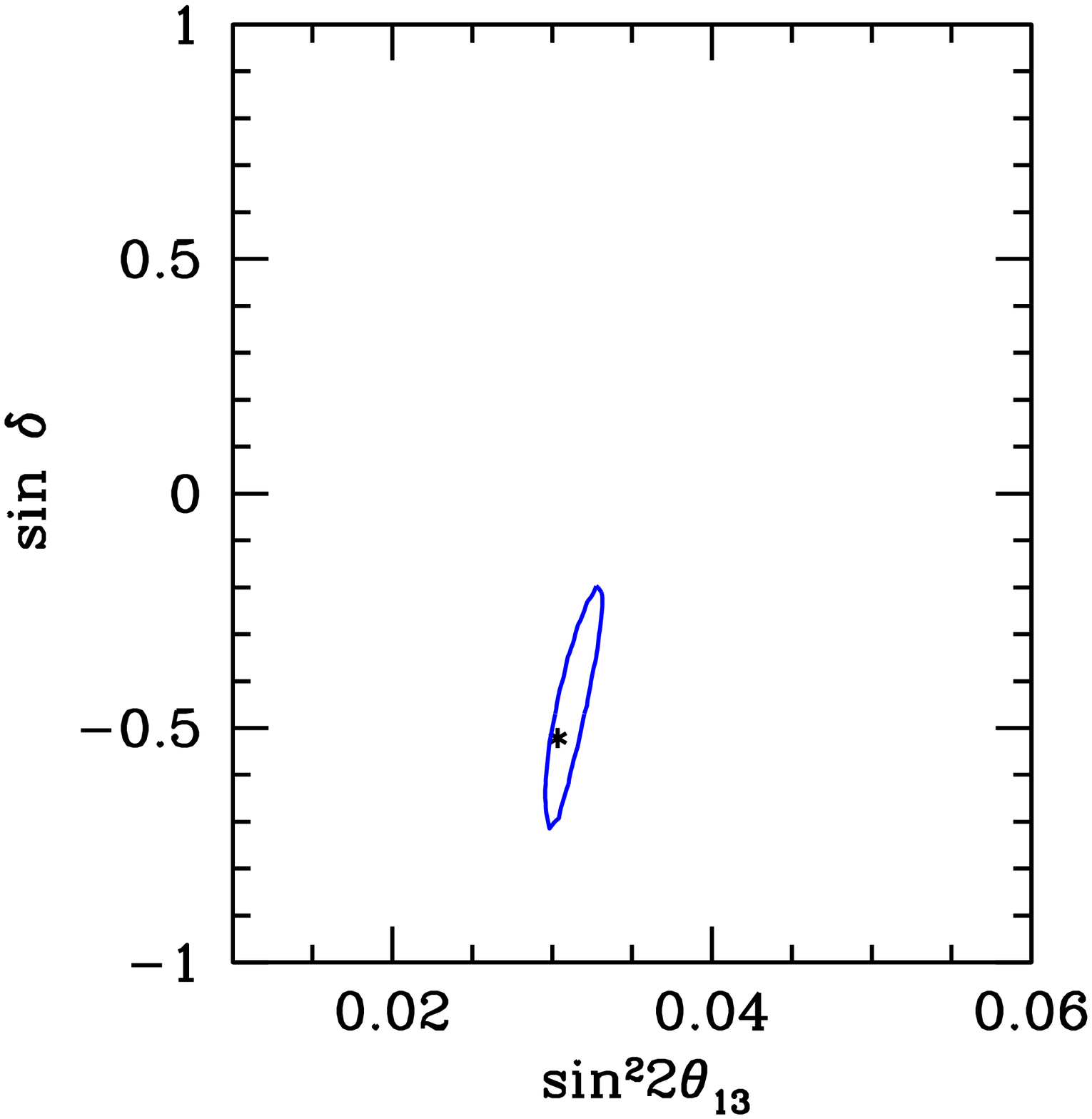, width=9.1cm} \\
\hskip 3.2truecm
{\small (a)}            &
\hskip 3.3truecm
{\small (b)}
\end{tabular}
\end{center}
\caption{\textit{(a) $95\%$ CL contours from a simultaneous $\chi^{2}$ fit to  
$\sin^{2} 2 \theta_{13}$ and $sin\delta$ without binning the data. The central value is denoted by a star: $\sin^{2} 2 \theta_{13} = 0.03$ and $\sin\delta=-0.5$. We have included in the former analysis statistical errors and backgrounds. The true (fake) solutions associated with the CP violating parameter $\sin \delta$ are depicted in solid (dotted) contours.
(b)The result of the combined $\chi^2$ analysis of the two energy bins.}}
\label{fig:fig1}
\end{figure}

Assuming a positive sign for the atmospheric mass splitting, the result of the fit assuming no energy binning in the signal for a particular central value  in the ($\sin^{2} 2 \theta_{13}, \sin \delta$) plane is shown in Fig.~(\ref{fig:fig1})(a), where two solutions arise, one of them (the fake one) corresponding to the wrong sign of $\cos \delta$. If the data analysis is performed with the opposite sign of the atmospheric mass splitting (negative), the two fake solutions associated to the wrong choice of the spectrum hierarchy are not present at the  $95\%$ CL for the particular central value chosen in Fig.~(\ref{fig:fig1}). 

\begin{figure}[h]
\begin{center}
\epsfig{file=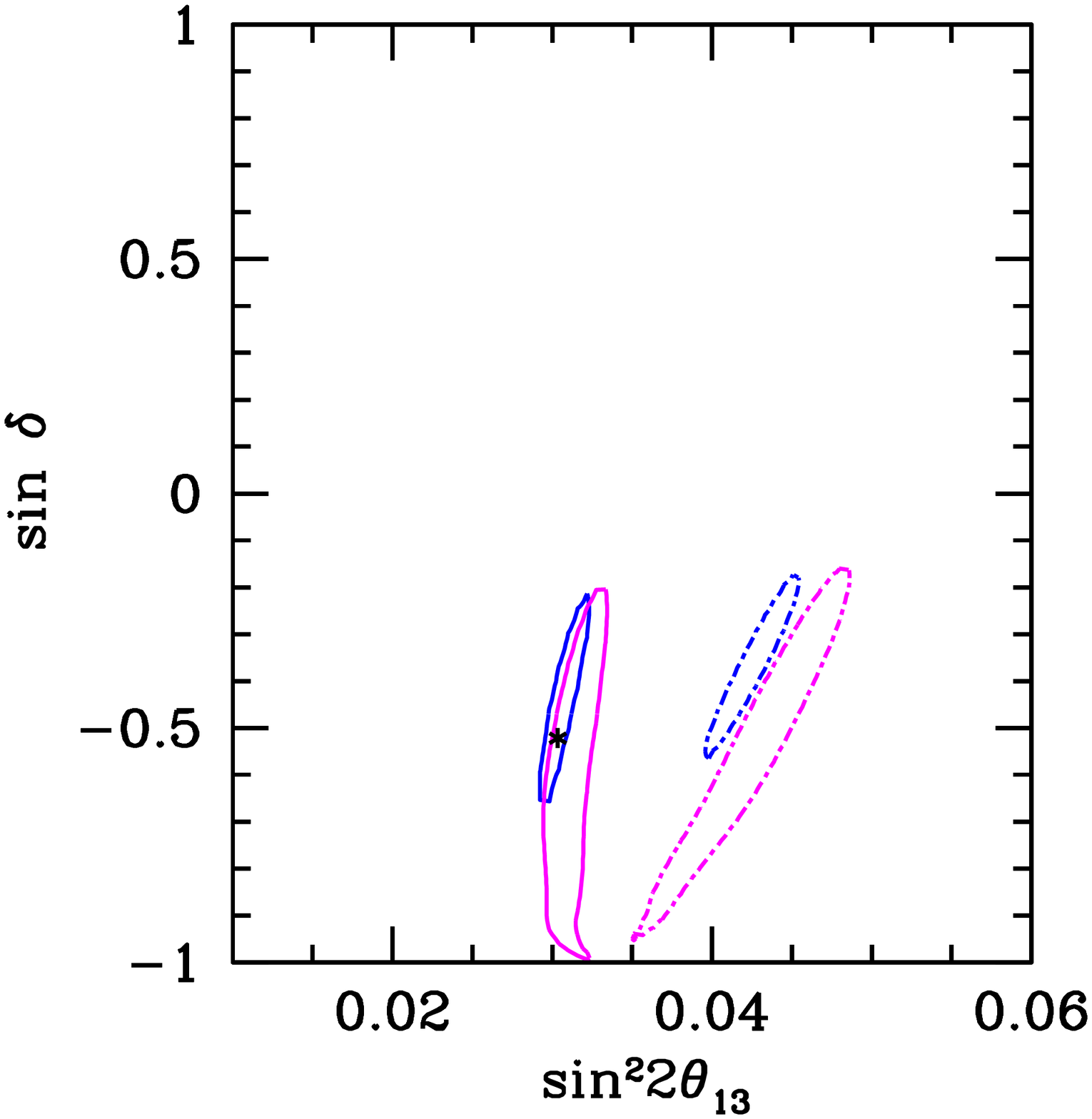, width=10.1cm,height=9.cm} 
\caption{\textit{$95\%$ CL contours from a simultaneous $\chi^{2}$ fit to  $\sin^{2} 2 \theta_{13}$ and $sin\delta$ applying a binning to the data. The results of the analysis with the first (second) bin of data are depicted in blue (magenta). We have included in the former analysis statistical errors and backgrounds. The fake solutions disappear when the two bins of data are combined, as shown in Fig.~(\ref{fig:fig1})(b).}}
\label{fig:fig2}
\end{center}
\end{figure}

If one exploits the energy information in the signal by performing an analysis in energy bins, the \emph{intrinsic, fake} solution is resolved, as it is shown in Fig.~(\ref{fig:fig1})(b). It has been shown that, for sufficiently large $\theta_{13}$ and in the vacuum approximation, apart of the true solution, there is a fake one at~\cite{deg1}:
\begin{eqnarray}
\sin\delta^{'}&\simeq&\sin \delta~,\nonumber\\
\theta^{'}_{13}&\simeq & \theta_{13} +
\cos\delta\ \sin 2\theta_{12} \;\Delta_{21} \cot \theta_{23} \cot\left(\Delta_{31}\right) \;.
\end{eqnarray}
The fake solution would be located at a value of $\sin^{2} 2 \theta_{13}^{'}$ which is energy-dependent. The degeneracies associated with the different energy bins would therefore have different locations in $\sin^{2} 2 \theta_{13}^{'}$. We illustrate the results for the two bins separately in Fig.~(\ref{fig:fig2}). The fake solutions appear in two different regions of the parameter space: when combining the information from the two separate bins, the intrinsic degeneracy would disappear. In order for this conclusion to hold true, the energy dependence of the signal has to be significant enough; otherwise, the analysis in energy bins would not provide an effective elimination of the fake solutions. We have performed the analysis in energy bins for smaller values of $\sin^2 2 \theta_{13}$ than the one shown in  Figs.~(\ref{fig:fig1}) and (\ref{fig:fig2}). We find that the intrinsic degeneracy is resolved if $\sin^2 2 \theta_{13}>0.02$.

\section{Conclusions}
We show the neutrino oscillation physics potential that can be achieved with the Fermilab NuMI beamline 10 km off-axis and three different experimental setups, differing in the proton luminosities and/or in the detector sizes. We provide a complete study of the sensitivities to $\sin^2 2\theta_{13}$, to the hierarchy of the neutrino mass spectrum, and to the CP violating parameter $\sin \delta$ for the three different scenarios.
We present our results in the ($\sin^{2} 2 \theta_{13}, \sin \delta$) plane; this choice helps enormously in understanding the location of the solutions for different experiments, and turns out to be very easy to generalize if $\theta_{23}\neq\pi/4$. We also explore the benefits of a modest energy resolution: with a $50\%$ energy resolution, the intrinsic degeneracies are  lifted if $\sin^{2} 2 \theta_{13} > 0.02$.

\section{Acknowledgments}
The authors would like to thank Adam Para for enlightening discussions and useful comments about the manuscript. Our calculations made extensive use of the Fermilab General-Purpose Computing Farms~\cite{Albert:2003vv}.
Fermilab is operated by Universities Research Association Inc.\ under
Contract No.\ DE-AC02-76CH03000 with the U.S.\ Department of Energy.  
\clearpage
\section{Appendix}
\subsection{Oscillated statistics} 
In Table~(\ref{tab:fluxes}) we provide the computed charged-current event rates at the NO$\nu$A far site (810 km) in the \textbf{Medium} experimental setup described in the Introductory Section.
\begin{table}[h]
\vspace{.5cm}
\centering
\begin{tabular}{|c|c|c|c|}
\hline\hline
$\nu_e$ (signal) & $\nu_e$ (background)& $\bar{\nu}_e$ (background)&$\nu_e$ ($\theta_{13}=0$)\\ \hline\hline
145 & 50.0 & 2.87 &7.55\\
\hline\hline 
$\bar{\nu}_e$ (signal) & $\nu_e$ (background)& $\bar{\nu}_e$ (background)&$\bar{\nu}_e$ ($\theta_{13}=0$)\\ 
\hline\hline
44.8 & 6.64 & 17.4 &2.33\\
\hline\hline 
\end{tabular}
\caption{\it 
Calculated charged currents neutrino and antineutrino event rates (signal and backgrounds) for NuMI (baseline of 810 km, 10 km off-axis) in the \textbf{Medium} experimental setup described in the Introduction. In order to compute the signal we have assumed energy independent oscillation probabilities P ($\nu_\mu\to \nu_e$) and P ($\bar{\nu}_\mu\to \bar{\nu}_e$) equal to $1\%$. The $\nu_e$ ($\theta_{13}=0$) and $\bar{\nu}_e$ ($\theta_{13}=0$) are the contributions from $P_\odot$ in Eq.~(\ref{eqn:e_appear}). The non oscillated $\nu_\mu$ ($\bar\nu_\mu$) event rate can be computed by multiplying the $\nu_e$ ( $\bar{\nu}_e$) signal by 100 and dividing the result by the $\nu_e$ ( $\bar{\nu}_e$) detection efficiency.}  
\label{tab:fluxes}
\end{table}
\newpage
\subsection{$\sin^2 2 \theta_{13}$ sensitivity} 
\begin{table}[h]
\centering
\begin{tabular}{||c|c||c|c|c|c||}
\hline\hline
&\textbf{mode/Setup}&\textbf{Small}&\textbf{Medium}&\textbf{Large}\\
\hline\hline
$\sin^2 2\theta_{13} $
&$\nu$ &$ 2.0\times 10^{-2}$&$1.6\times 10^{-2}$&$1.0 \times 10^{-2}$\\
\cline{2-5}
$99\% $ CL-sensitivity
&$\bar\nu$ &$4.0\times 10^{-2}$&$2.0\times 10^{-2}$ & $1.3\times 10^{-2}$\\
\cline{2-5} 
(most restrictive)
&$\nu+\bar\nu$ &$1.2\times 10^{-2}$&$ 7.0\times 10^{-3}$&$4.0\times 10^{-3}$\\
\hline\hline
$\sin^2 2\theta_{13}$ 
&$\nu$ &$4.0\times 10^{-3}$&$ 1.2\times 10^{-3}$& $3.0\times 10^{-4}$\\
\cline{2-5}
$99\% $ CL-discovery
&$\bar\nu$ &$1.0\times 10^{-2}$&$4.0\times 10^{-3}$&$1.5\times 10^{-3}$ \\
\cline{2-5}
(least restrictive)
&$\nu+\bar\nu$ &$4.0\times 10^{-3}$&$ 1.2\times 10^{-3}$& $3.0\times 10^{-4}$\\
\hline\hline
\end{tabular}
\caption{\it{$99\%$ CL sensitivity limits to $\sin^2 2 \theta_{13}$, extracted from Figs.~(\ref{fig:sensnu}), (\ref{fig:sensanu}) and (\ref{fig:sensnuanu}), using only the neutrino channel, only the antineutrino one, or both channels, respectively. The first three rows indicate the most restrictive limits which is independent of the value of the CP phase $\delta$ and of the type of hierarchy chosen by nature. The last three rows show the $99\%$ CL-$\sin^2 2 \theta_{13}$ discovery limits for the most favorable choice of the parameters. The three columns refer to the three different experimental setups explored here.}} 
\label{tab:th13sens}
\end{table}

\end{document}